\documentclass{PoS}

\title{CTEQ-TEA group updates: Photon PDF and Impact from heavy flavors in the CT18 global analysis}
\ShortTitle{The photon and heavy flavors in the CT18 PDFs}

\author{
Marco Guzzi$^1$\thanks{mguzzi@kennesaw.edu},
\speaker{Keping Xie$^2$}\thanks{xiekeping@pitt.edu},
Tie-Jiun Hou$^3$,
Pavel Nadolsky$^4$,
Carl Schmidt$^5$,
Mengshi Yan$^6$,
and C.-P. Yuan$^5$
\\
$^1$Department of Physics, Kennesaw State University, Kennesaw, GA 30144, USA\\
$^2$Department of Physics and Astronomy, University of Pittsburgh, Pittsburgh, PA 15260, USA\\
$^3$Department of Physics, College of Sciences, Northeastern University, Shenyang 110819, China\\
$^4$Department of Physics, Southern Methodist University, Dallas, TX 75275, USA\\
$^5$Department of Physics and Astronomy, Michigan State University, East Lansing, MI 48824,USA\\
$^6$School of Physics and State Key Laboratory of Nuclear Physics and Technology,
Peking University, Beijing 100871, China
}

\abstract{
We discuss recent CTEQ-TEA group activities after the publication of the
CT18 global analysis of parton distribution functions (PDFs) in the proton.
In particular, we discuss a new calculation for the photon content in the
proton, termed as CT18lux and CT18qed PDFs,
and the impact of novel charm- and bottom-quark production cross section
measurements at HERA on the CT18 global analysis.

\hfill{MSUHEP-21-028, PITT-PACC-2121, SMU-HEP-21-14}
}

\FullConference{The European Physical Society Conference on High Energy Physics (EPS-HEP) 2021\\
July 26-30, 2021\\
Online Conference}

\begin{document}

{\bf Introduction.}
The large inflow of high-precision measurements from the Large Hadron Collider (LHC) brings hadron collider phenomenology in a new era in which the LHC is a precision machine.
To match the accuracy and precision of experimental data, remarkable progress has been made in the calculation of theory predictions for standard candle cross sections which now include radiative corrections at next-to-next-to-leading order (NNLO) ${\cal O}(\alpha_s^2)$ and beyond in the QCD strong coupling $\alpha_s$, and at next-to-leading order (NLO) ${\cal O}(\alpha_{EW})$ in the electroweak (EW) coupling for numerous processes.
At such level of precision, there are several effects whose magnitude can compete in size with that of radiative corrections at NNLO, or N$^3$LO, and that have observable impacts.
Examples of these effects, which we shall discuss below, are: (1) consistent inclusion of EW corrections in the initial state of parton reactions in proton-proton collisions, (2) phase-space suppression and other mass effects that are comparable in magnitude to higher-order corrections in QCD and EW.
These effects have been recently investigated by the CTEQ-TEA group~\cite{Xie:2021equ,Guzzi:2021gvv,Guzzi:2021fre,Anchordoqui:2021ghd} after the publication of the CT18 global analysis~\cite{Hou:2019efy} of parton distribution functions (PDFs) of the proton.

{\bf Impact of photon PDF in the CT18 global analysis.}
The photon PDF consists of a large elastic contribution where
the proton remains intact, and of an inelastic contribution in
which the proton breaks into a multihadron final state cf.~\cite{Martin:2014nqa}.
According to this decomposition, various methods have been introduced in literature to parametrize the photon content in the proton. These can be grouped in first and second generation models.
First generation models include the MRST2004QED~\cite{Martin:2004dh}, NNPDF2.3QED~\cite{Ball:2013hta}, NNPDF3.0QED~\cite{Ball:2014uwa}, and CT14QED~\cite{Schmidt:2015zda} analyses.
In the CT14QED analysis, the inelastic contribution
to the photon PDF was described by a two-parameter ansatz, coming from radiation off
the valence quarks, based on the CT14 NLO PDFs~\cite{Dulat:2015mca},
and constrained by using isolated photon production ($e p\rightarrow e\gamma + X$) cross section measurements~\cite{Chekanov:2009dq} in deep inelastic scattering (DIS) from the ZEUS collaboration. The elastic contribution is included in the CT14QEDinc PDFs where the inclusive photon PDF at $Q_0$ is defined by the sum of the inelastic and elastic contributions obtained from
the Equivalent Photon Approximation (EPA)~\cite{Budnev:1974de}.
The second generation models are described in the LUXqed~\cite{Manohar:2016nzj,Manohar:2017eqh}, NNPDF3.1luxQED~\cite{Bertone:2017bme}, MMHT2015qed~\cite{Harland-Lang:2019pla}, and CT18qed~\cite{Xie:2021equ} analyses.
The CT18qed study aims at constraining the photon PDF by using the LUXqed ansatz with two approaches. In one approach, the CT18lux fit, the photon PDF is calculated directly using the LUXqed master formula at any scale $\mu$. In this case, quark and gluon PDFs are inherited from the CT18 NNLO
global fit with no modifications. The momentum sum rule of CT18lux PDFs is weakly violated as the photon enters as an additional small component (about $0.2\%$ at 1.3 GeV).
In an alternative realization, the CT18qed, the photon PDF is initialized using the LUXqed formula at a lower scale, $\mu\sim\mu_0$, and evolved to higher scales with a combined QED kernel at $\mathcal{O}(\alpha)$, $\mathcal{O}(\alpha\alpha_s)$ and $\mathcal{O}(\alpha^2)$.
The difference between CT18lux and CT18qed is mainly ascribed to the size of missing higher-order contributions when different scales are matched.
\begin{figure}[ht!]\hspace{-1.2cm}
	\includegraphics[width=0.52\textwidth]{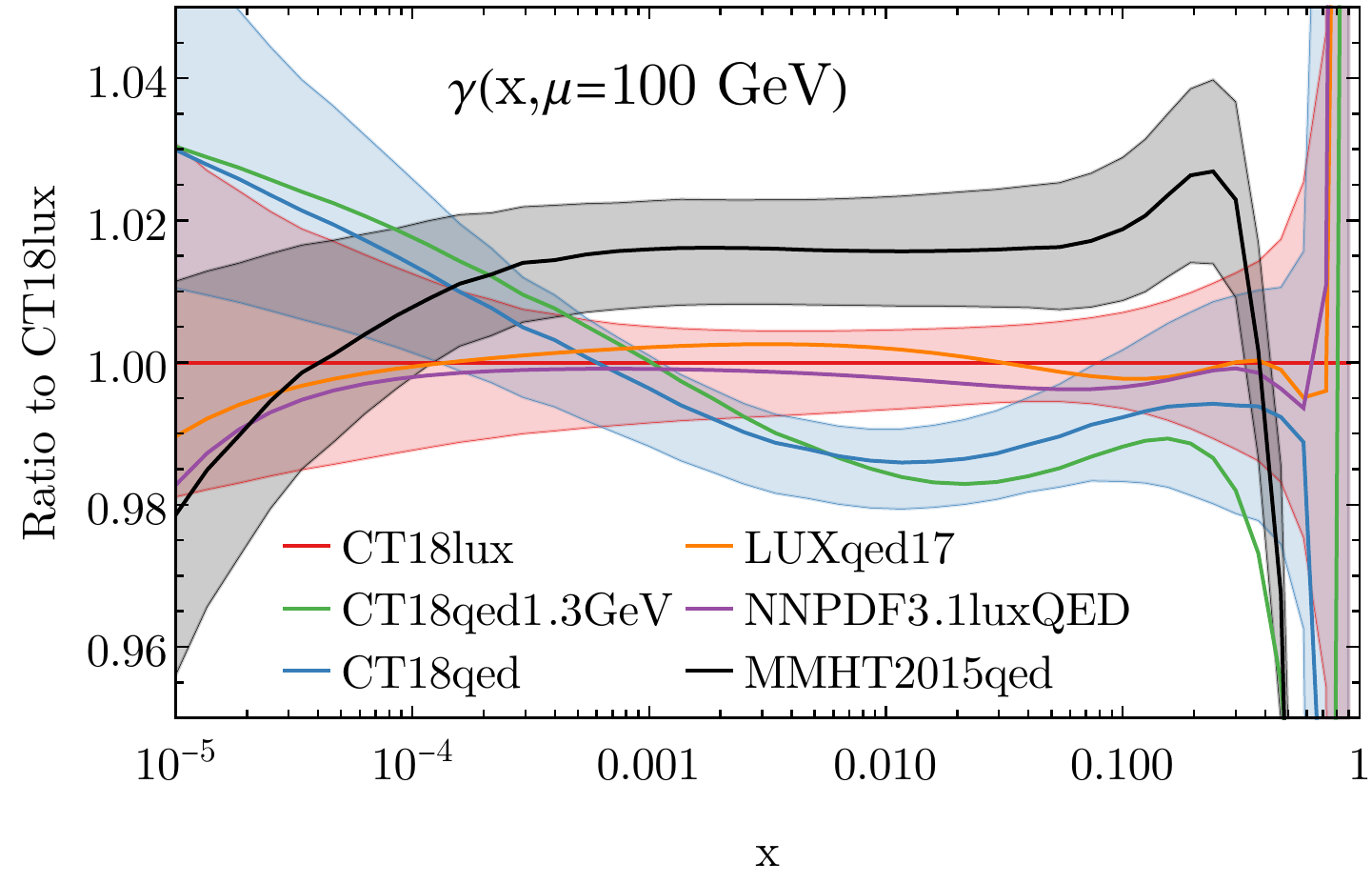}
	\includegraphics[width=0.52\textwidth]{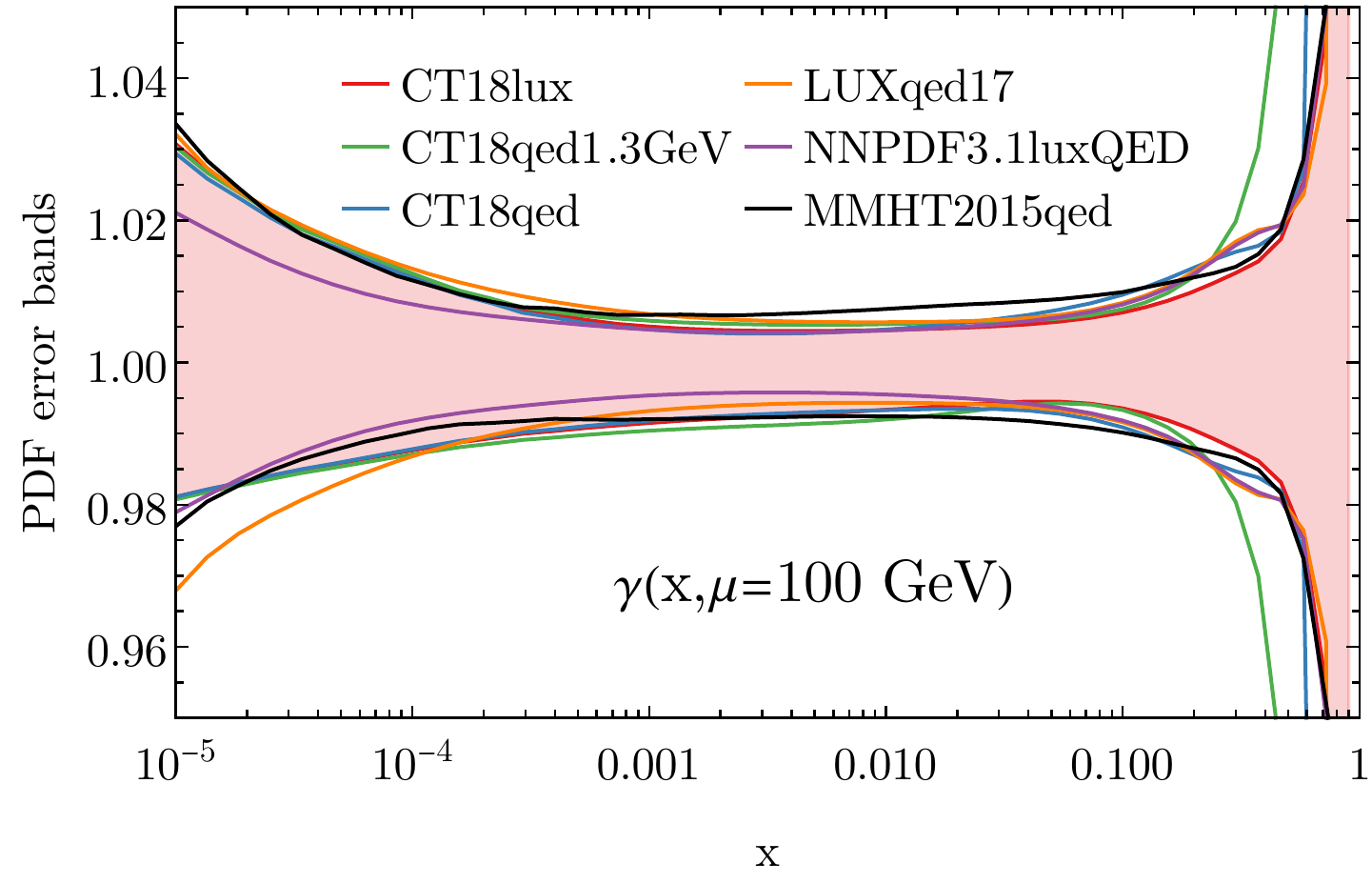}
	\caption{
	Left: A comparison of the photon PDFs among the CT18lux, CT18qed(1.3GeV), LUXqed17 \cite{Manohar:2017eqh}, NNPDF3.1luxQED \cite{Bertone:2017bme}, and MMHT15qed \cite{Harland-Lang:2019pla}.
	Right: Plot of the self-normalized uncertainty bands for each of the photon PDFs examined in this analysis.
	}
	\label{fig1}
\end{figure}
It is important to point out that integrals of the unpolarized electromagnetic
structure functions (SFs) $F_{2,L}(x,Q^2)$ in the LUX formalism are carried out over a wide rage of $Q^2$ values which can be sensitive to higher twists and other nonperturbative QCD contributions. These require explicit modeling as they are not suppressed at low $Q^2$ and must be accounted for to get a genuine estimate of the uncertainties.
On the other hand, in a typical global PDF analysis at NNLO, only leading-twist contributions are considered in the factorization formula where the $Q^2$ and $W^2$ values are chosen such that  sub-leading twist effects are suppressed.
In Figure~\ref{fig1}, we illustrate a comparison between the central values and uncertainties for the photon PDF at $\mu=100$ GeV as obtained by different groups. This summarizes the main findings of the CT18qed analysis. In the right inset, we notice that in the intermediate-$x$ region ($10^{-3}\leq x\leq 0.1$), all photon PDFs exhibit comparable error bands. In the left inset, the CT18lux photon PDF lies in between that of LUXqed (and also NNPDF3.1luxQED) and MMHT2015qed, while the CT18qed photon appears to be suppressed. In the small-$x$ region, the CT18qed photon is larger than that of CT18lux due to the $x$ behavior of the SF ratio $F_2^{LO}/F_2^{NNLO}$ at high scale $\mu$ (cf.~\cite{Xie:2021equ} and related discussion). The MMHT2015qed photon instead becomes smaller because the singlet PDF combination becomes smaller.
In the large-$x$ region, both MMHT2015qed and CT18qed photon PDFs are smaller as compared to that obtained in the LUX approach. This is because the inelastic contribution to the photon PDF receives substantial nonperturbative corrections through the structure functions $F_{2,L}$ at the initial lower scale $\mu_0$. Therefore, DGLAP evolution in both MMHT2015qed and CT18qed results in smaller PDFs. However, the nonperturbative contributions to the structure functions at low scales in the DGLAP approach result in larger uncertainty at large $x$~\cite{Xie:2021equ}.
The CT18qed PDFs are the recommended CTEQ-TEA PDFs with the photon included as an active parton inside the proton, and can be used for theory predictions at any scale $\mu$ greater than 3 GeV.\footnote{As compared to the CT18 analysis which uses a starting scale $\mu_0=1.3$ GeV, here we choose as default scale the PDF matching scale $\mu_0=Q_{\rm PDF}=3$ GeV. This reduces the uncertainty resulting from low-scale nonperturbative contributions in the structure functions~\cite{Xie:2021equ}.}

{\bf The impact of heavy-flavor production on the CT18 PDFs.}
Other contributions which compete in size with higher-order corrections in QCD and EW are related to mass effects and phase-space suppression due to heavy-flavor production in DIS reactions. DIS cross section measurements constitute the most important component of data ensembles in global analyses of PDFs. Therefore, it is critical to consistently account for the heavy-parton mass effects in the factorization formula.
The modern literature~\cite{Aivazis:1993kh,Aivazis:1993pi,Buza:1996wv,Thorne:1997ga,Kramer:2000hn,Tung:2001mv,Forte:2010ta,Alekhin:2009ni,Guzzi:2011ew}
~contains various modifications of the factorization theorem that are introduced to study heavy-flavor production in DIS structure functions and that are currently employed in recent global QCD analyses to determine proton's PDFs~\cite{Ball:2021leu,Bailey:2020ooq,Hou:2019efy,NNPDF:2017mvq,Alekhin:2017kpj}.
The neutral current (NC) DIS data of charm and bottom production for the structure functions $F_2^{c \bar{c}}$ and $F_2^{b \bar{b}}$ at high $Q^2$~\cite{H1:2004esl}, and the reduced cross sections $\sigma_{red}^{c \bar{c}}$ for the combined charm data at HERA~\cite{H1:2012xnw} provided important information about proton's PDFs, in particular the gluon and strange-quark PDFs, in the CT18 global analysis~\cite{Hou:2019efy}.
These data~\cite{H1:2004esl,H1:2012xnw}, were recently updated by the H1 and ZEUS collaborations which published a new set of charm and bottom production measurements~\cite{H1:2018flt} in an extended kinematic region, and with much smaller uncertainties due to simultaneous combination of $c$ and $b$ data.
When the new charm and bottom combination~\cite{H1:2018flt} is included in the CT18 global analysis at NNLO, the fit of these data becomes challenging. Our preliminary study shows that in the best scenario, the optimal description corresponds to a $\chi^2/N_{pt}$ that is no less than 1.7. In contrast,
in the CT18NNLO fit, we obtain $\chi^2/N_{pt}=1.98$ for charm production ($N_{pt}=47$), and $\chi^2/N_{pt}=1.25$ for bottom production ($N_{pt}=26$); while in the CT18XNNLO fit (a variant of CT18NNLO) we obtain $\chi^2/N_{pt}=1.71$ for charm and 1.26 for bottom production.
The new $c$, $b$ combination shows disagreement with several processes included in the CT18 baseline. Examples are $Z$ production at LHCb at 7 and 8 TeV, and CDF run-II,
and single inclusive jet production,
and $p_T$ and $y$ double differential cross sections in $t\bar{t}$ production at CMS 8 TeV.
For this reason, these data~\cite{H1:2018flt} were not included in the CT18 global analysis.
In this preliminary study, we investigated to a deeper extent features and impact of the
new $c$, and $b$ combined measurements~\cite{H1:2018flt} from H1 and ZEUS on the CT18 PDFs.
We explored the new correlated systematic uncertainties and performed dedicated studies of the parameters entering the theory calculation of the NNLO SFs.
The complete results of this investigation will be published in a forthcoming paper.

\begin{figure}[!ht]\hspace{-1.0cm}
\includegraphics[width=0.55\textwidth]{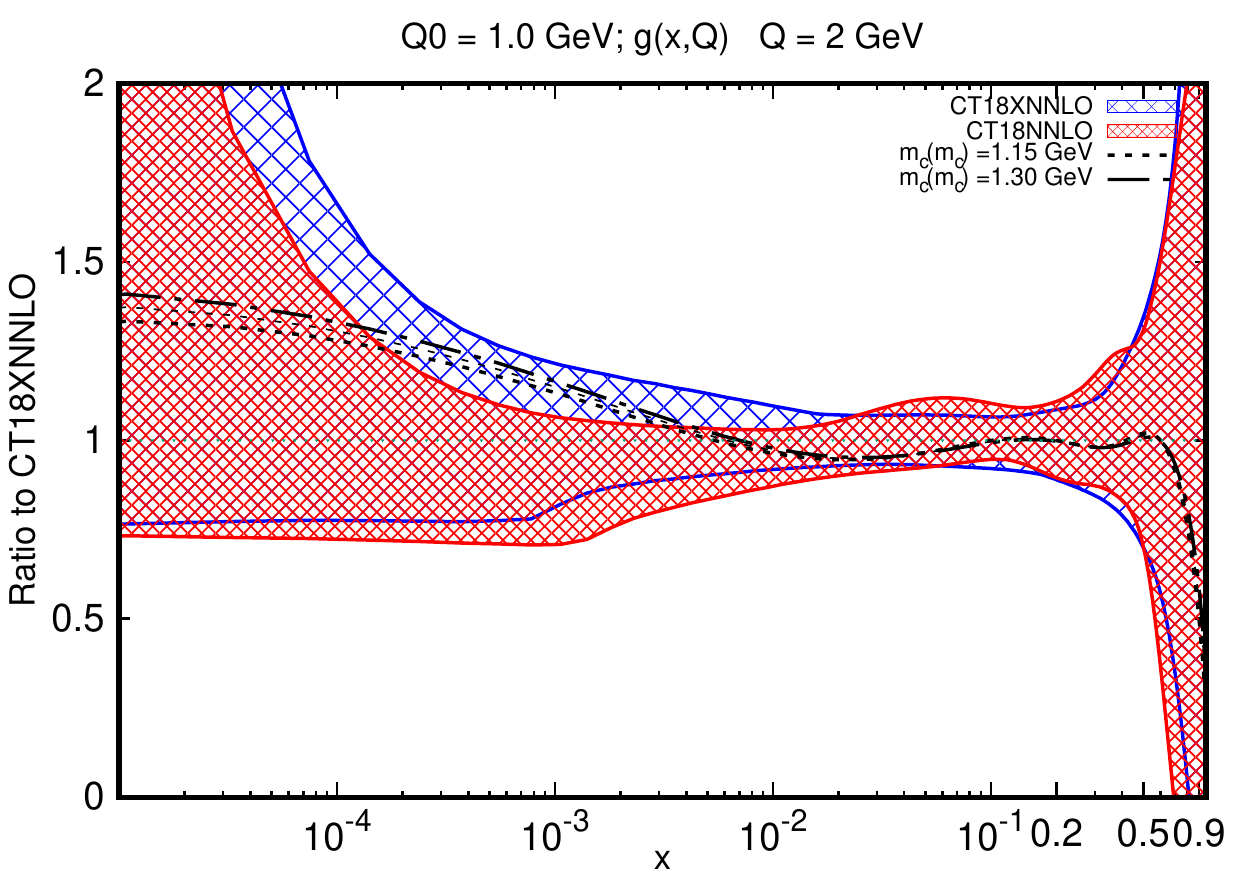}
\includegraphics[width=0.55\textwidth]{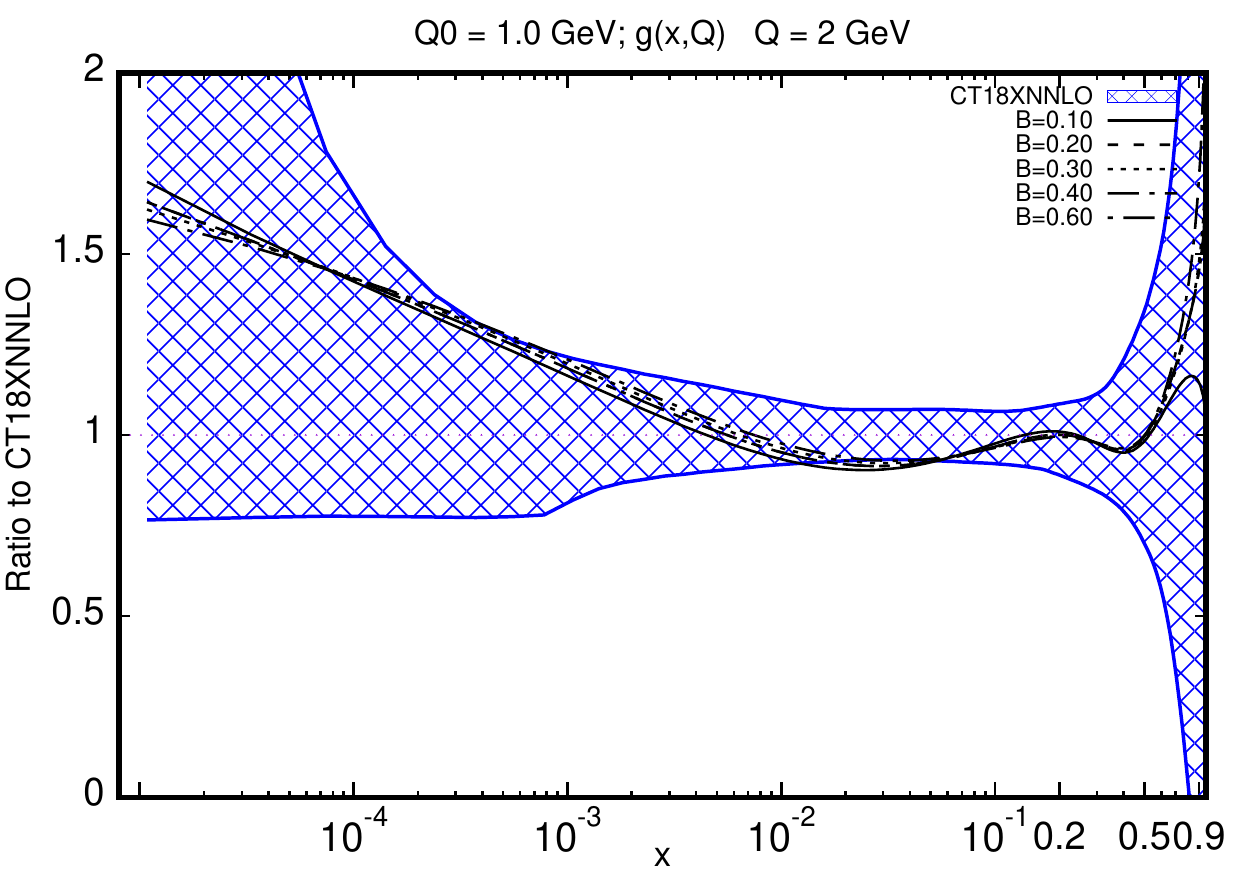}
\caption{Ratio to the CT18XNNLO gluon PDF as a function of $x$ at $Q=2$ GeV and $Q_0 = 1$ GeV.
Left: scan over the $\overline{\textrm{MS}}$ charm-quark mass $m_c(m_c)$ while $m_b(m_b)=4.18$ GeV. Right: scan over the $B$ parameter in  $\mu_{\textrm{DIS}}(x)$.
Error bands are shown at 90\% confidence level for CT18NNLO (red) and CT18XNNLO (blue).}
\label{fig:gluon}
\end{figure}

{\bf Description of the charm and bottom combination at HERA.}
As reported in several studies by different groups~\cite{H1:2015ubc,Alekhin:2017kpj,Thorne:2012az,dis2021-nocera,Bailey:2020ooq,Ball:2021leu} the $\chi^2$ description for the  new $c$ and $b$ combined measurements at HERA~\cite{H1:2018flt} is poor, and the theory seems to fail to describe the slope of the data in the intermediate/small $x$ region $10^{-5}\leq x\leq 0.01$.
To improve the description, we tried to vary different parameters in the fit, in combination with different fit settings.
Heavy flavors in DIS structure functions are treated according to S-ACOT-$\chi$ NNLO general mass variable flavor number (GMVFN) scheme~\cite{Guzzi:2011ew}, which is the default scheme utilized in all CTEQ analyses.
In one attempt to improve the description of these data we varied parameters of the $x$-dependent DIS factorization scale, defined as $\mu_{\textrm{DIS}}(x) = A\sqrt{m_Q^2+B^2/x^C}$,
and used for the calculation of low-$x$ DIS cross sections in the CT18XNNLO fit. The CT18XNNLO fit (a variant of CT18NNLO) is generated by including the $\mu_{\textrm{DIS}}(x)$ scale choice for low-$x$ DIS data.
This $x$-dependent scale choice is expected to mimic the main behavior of low-$x$ resummation~\cite{Ball:2017otu} and is inspired by saturation models \cite{Golec-Biernat:1998zce,Caola:2009iy}.
In Fig.~\ref{fig:gluon}, we show the changes on the gluon PDF at $Q=2$ GeV and $Q_0 = 1$ GeV at NNLO when two parameters, the $\overline{\textrm{MS}}$ charm-quark mass $m_c(m_c)$, and $B$, are varied in $\mu_{\textrm{DIS}}(x)$, after fixing the parameters $A=0.5$ and $C=0.33$.
In the left inset, the solid black curve corresponds to the fit with $m_c(m_c)=1.15$ GeV with $\chi^2(\textrm{HERA HQ})/N_{pt}=1.62$,
while the dotdashed represents the fit with $m_c(m_c)=1.30$ GeV and with $\chi^2(\textrm{HERA HQ})/N_{pt}=2.33$.
In the right inset, the solid black curve corresponds to the fit with $B=0.10$ GeV and $\chi^2(\textrm{HERA HQ})/N_{pt}=1.58$,
while the dotdashed represents the fit with $B=0.60$ GeV and $\chi^2(\textrm{HERA HQ})/N_{pt}=1.52$.  Error bands are shown at the 90\% confidence level for CT18NNLO and CT18XNNLO. Overall, these preliminary findings indicate that the new charm and bottom production measurements at HERA seem to have a preference for a harder gluon at intermediate and small $x$.

{\bf Acknowledgements.}
We thank the CTEQ-TEA group for helpful discussions.
The work of M. Guzzi is supported by the U.S.~National Science Foundation under Grant No.~PHY~2112025.
The work of C.-P.~Yuan is partially supported by the U.S.~National Science Foundation under Grant No.~PHY-2013791.
The work of K.~Xie was supported in part by the U.S.~Department of Energy under
grant No.~DE-FG02-95ER40896, U.S.~National Science Foundation under Grant No.~PHY-1820760, and in part by the PITT PACC.
The work of M.~Yan is supported by the National Science Foundation of China under Grant Nos. 11725520, 11675002, and 11635001.
C.-P.~Yuan is also grateful for the support from the Wu-Ki Tung endowed chair in particle physics.

\bibliographystyle{utphys}
\providecommand{\href}[2]{#2}\begingroup\raggedright\endgroup

\end{document}